# Preventive and Active Safety Applications

**Levent Güvenç**
Automotive Controls Research Group
Department of Mechanical Engineering
İstanbul Technical University
Gümüşsuyu, Taksim, TR-34437, İstanbul, Turkey
Fax: +90 212 245 07 95, E-mail: guvencl@itu.edu.tr

**ABSTRACT**

Road vehicle safety systems can be broadly classified into the two categories of passive and active systems. The aim of passive safety systems is to reduce risk of injury to the occupants of the vehicle during and after an accident like a crash or rollover. Passive safety systems include the design of safety restraints, design for crashworthiness, seat belts and air bags. In contrast to passive systems, the aim in active safety is to prevent an accident from occuring in the first place. As such, it makes sense to call them preventive systems also. This paper concentrates on preventive and active safety systems. The current state of the art in some key preventive and active safety systems is presented in this paper, wherein the various techniques used are also explained briefly. In some cases, the presentation is complemented with results obtained in the author's research group. A road map of expected future developments in the area of preventive and safety applications is also presented.

**INTRODUCTION**

Road vehicle safety systems can be broadly classified into passive and active safety systems. Passive safety systems include seat belts, air bags and additional structural members. The objective of passive safety systems is to reduce damage to the driver and passengers of a road vehicle during and after an accident like a crash or rollover. Active safety systems work before an accident and aim at preventing an accident from happening in the first place. As such, it makes sense to call them preventive systems also. This prevention is achieved by trying to keep the vehicle from exhibiting undesired motions like wheel lock-up, loss of traction or excessive yaw or roll motion that may eventually result in loss of control of the vehicle's dynamic behaviour by the driver. Active safety systems may, thus, take control away from the driver temporarily, until the undesired vehicle dynamic behavior is corrected.

In Europe, 160 billion Euros is lost in accidents. This is equivalent to 2% of the GNP of Europe. There are 41,000 deaths and a very large number of injuries as a result of these road vehicle accidents, per year. The European Union has started the eSafety initiative through its FP6 programs with an aim of reducing road vehicle accident fatalities by 50% by the year 2010. Preventive and active safety applications will be a vital part of such an effort.

The current state of the art in preventive and active safety systems are discussed in this paper. First, more commonly available systems like ABS, TCS and ESP are discussed. Rollover threat warning and rollover avoidance systems are discussed next. Lane keeping, intelligent transportation, collision avoidance and adaptive cruise control systems are discussed afterwards. The use of hardware-in-the-loop (HiL) and simulator tests to develop active safety systems is discussed. The paper ends with a brief presentation of active safety control system work conducted by the Automotive Controls Research Group in the Department of Mechanical Engineering of İstanbul Technical University and a road map of expected future developments in this area along with recommendations and conclusions.

**ABS BRAKES**

Anti-lock brakes (ABS) are the oldest and most successful active safety systems in road vehicles. A typical ABS system consists of a hydraulic modulator, a hydraulic power source, wheel speed sensors and an electronic control unit. ABS systems sense impending wheel lock-up and prevent it by reducing brake pressure. ABS systems can be viewed as wheel longitudinal slip controllers and are almost a standard in present day vehicles. Proper ABS system design considering all possible ranges of loading, road conditions and vehicle speed is still an important challenge. This challenge is more demanding for commercial vehicles. New developments in this area will be adaptation of ABS brake technology to brake-by-wire systems and to hybrid electric vehicles with electric motor powered traction.

**TCS SYSTEM**

Traction control systems can also be viewed as longitudinal slip controllers like ABS brakes but they operate after encountering large values of longitudinal slip that occur when (one) wheel(s) spin without traction on the road surface. This happens quite commonly on startup on icy or muddy road or a sudden acceleration request during normal driving. In such cases, a traction control system intervenes by applying the brake to the wheel that is spinning until traction is achieved. The result is close to optimum traction on all four wheels. TCS systems are also becoming standard components on new vehicles. The challenges and new developments for TCS are similar to those for ABS brakes presented in the section above. Future developments may also include the combination of ABS, TCS and ESP into one standard vehicle dynamics control unit.

**ESP**

Loss of yaw stability of a road vehicle may result from unexpected yaw disturbances like side wind force, tire pressure loss or µ-split braking due to unilaterally different road pavements such as icy, wet or dry pavement. An average driver may exhibit panic reaction and control authority failure and may not be able to generate adequate steering, braking/throttle commands in such panic situations. Road vehicle yaw stability control systems compensate for the driver's inadequacy during panic situations and generate the necessary corrective yaw moments through steering or braking control inputs. Such yaw stability control systems are called Electronic Stability Program (ESP).

The two primary corrective yaw moment generating methods of actuation for ESP systems are compensation using steering commands or using individual wheel braking. Most of the commercially available ESP systems use individual wheel braking as it is more easily accomplished through already existing ABS hardware (see for ex. van Zanten et al, 1995). Steering actuation is the second method of generating corrective yaw moments. Steering actuation can be in the form of a steer-by-wire system or through active steering. In active steering, the mechanical steering linkage is complemented by an extra steering motor which provides extra steering moment to the system. This is a fail safe approach as the mechanical steering system is in place in case of failure of the electric motor. Active steering can be used for implementing power steering, velocity dependent steering ratio or a yaw stability control system. The disadvantage in the case of a yaw stability control system is that the steering wheel will also move as the stability controller commands corrective steering signals. This is not a very good man-machine interface as the driver will definitely feel the unpleasant loss of his/her control of the vehicle when the yaw stability controller becomes active. A second disadvantage will be a slight loss of responsiveness as the whole steering linkage including the steering wheel has to be moved by the electric motor used for control. Active steering has been available in some cars for a while.

In contrast to active steering, steer-by-wire systems offer more flexibility for yaw stability controller implementation as the full steering command is available to the controller. Steer-by-wire systems possess only electrical connections between the steering wheel and the steering actuator. This offers great flexibility and solves several problems at the expense of not having a mechanical backup system. Steer-by-wire systems are also available for implementation in production vehicles. This technology enables easy implementation of steering based ESP systems (see for example Ackermann et al, 2002; Aksun Güvenç and Güvenç, 2002a).

The biggest overall gain is achieved by combining both steering and braking actuation for more corrective yaw moment when necessary. This methodology is similar to what drivers actually do during their panic reaction. The simultaneous use mentioned has to be performed in a coordinated manner. Combined and coordinated use also allows the control system to switch between actuation methods in the event of an actuator malfunction. In related work, Aksun Güvenç and Güvenç (2002a, 2002b) have developed a very useful robust steering based ESP system using a modified model regulator. Realistic simulations and actual road tests of this modified model regulator approach have been successful. Work on an individual wheel braking type implementation of this proposed control law have been conducted by the same authors (Aksun Güvenç and Güvenç, 2002c; Aksun Güvenç et al., 2003a). Previous work on coordinated use of both means of actuation has been presented in Aksun Güvenç et al (2003b, 2004). Recent work on the use of this combined actuation ESP controller in a unified framework that allows its use as a steering only, braking only or combined actuation setting and to demonstrate its usefulness on a realistic, high-fidelity vehicle model in contrast to the simple, lower order models used in previous work has been presented in Öztürk et al (2005).

Some sample simulation results taken from Öztürk (2004) are shown in Figures 1 and 2. Both simulations used a typical SUV modeled in Adams®/Car as the plant while the combined actuation controller mentioned above was implemented as the controller in Matlab®/Simulink®. Co-simulation of both programs was used to obtain the results. A double lane change maneuver under the action of side wind force is shown in Figure 1. The dark SUV with the combined actuation ESP controller has no problem in performing the desired double lane change. The light colored SUV in the same figure corresponds to the conventional SUV without an ESP and strays off course under the disturbing action of side wind. A µ-split maneuver is shown in Figure 2. The light shaded area represents an ice patch. The rest of the road is dry. As the vehicle enters the ice patch with its right hand set of tires, the brakes are applied fully. The difference in friction coefficients on the left and right hand sides results in a yaw moment that makes the uncontrolled, light colored SUV spin. In contrast, the dark colored SUV with the combined ESP controller is able to generate the necessary corrective yaw moments and keep its lateral stability as seen in Figure 2.

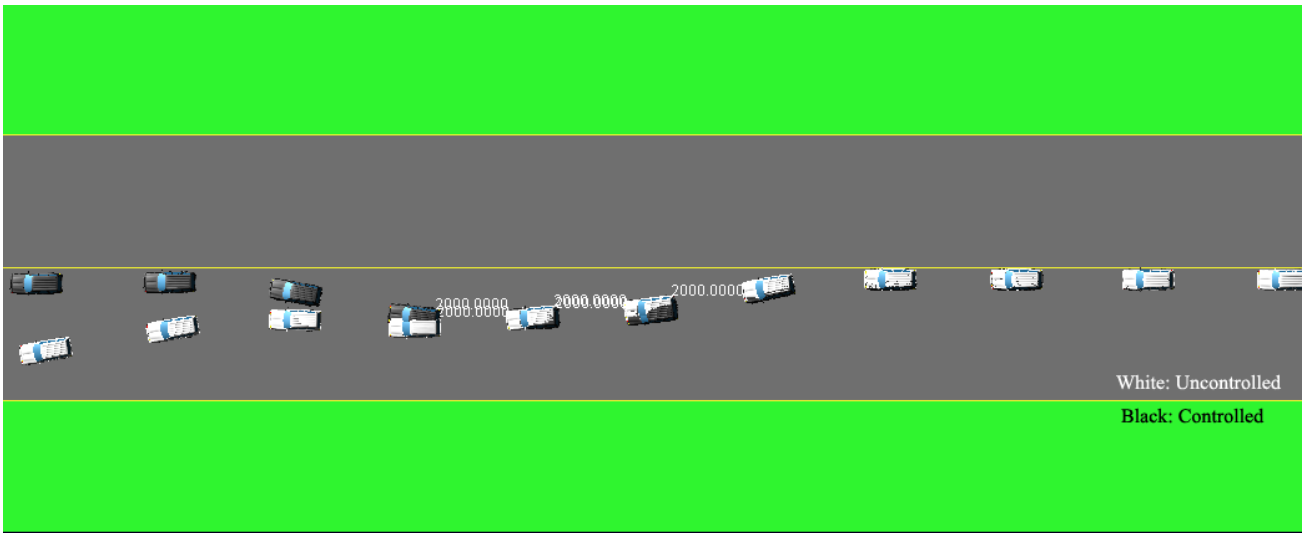

Figure 1 Effect of side wind on a lane change maneuver (Öztürk, 2004)

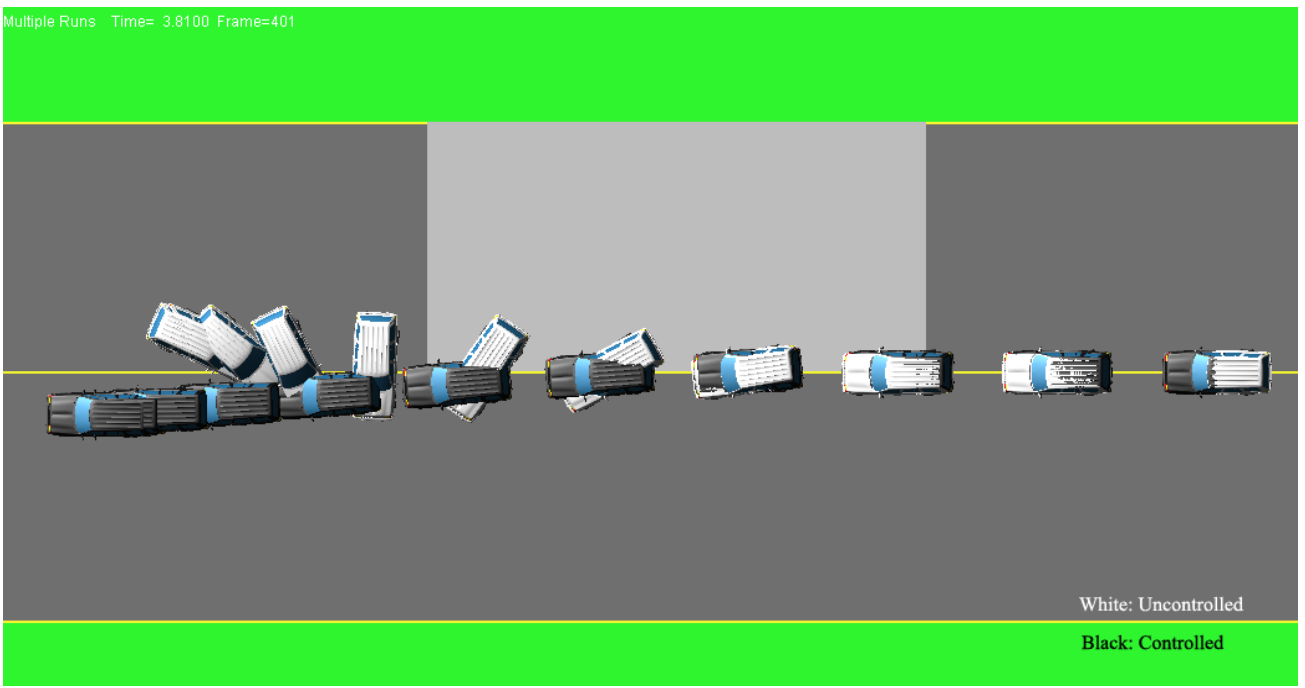

Figure 2 μ-split maneuver (Öztürk, 2004)

Even though it is not as common as ABS and TCS systems, ESP is also becoming popular with new car buyers. ESP is still sold as an extra option but it has become available for economy class cars also. In the future, ESP needs to enter the commercial vehicle market. One of the main obstacles hindering the use of ESP for commercial vehicles is that the commercial vehicle user does not want to pay the extra cost. The large variations in mass properties and center of gravity location also make it difficult to tune ESP controllers for commercial vehicles.

A projection into the future of ESP systems will begin with the use of brake-by-wire and steer-by-wire systems in production vehicles and the associated changes in the ESP controller. The use of by-wire technology will require more work in the future on fail safe design and fault tolerance. As hybrid vehicles are becoming more widespread, ESP systems that use the advantages of electric motor based traction and braking (if applicable) should be developed. There has also been considerable successful research effort at using GPS information in ESP systems.

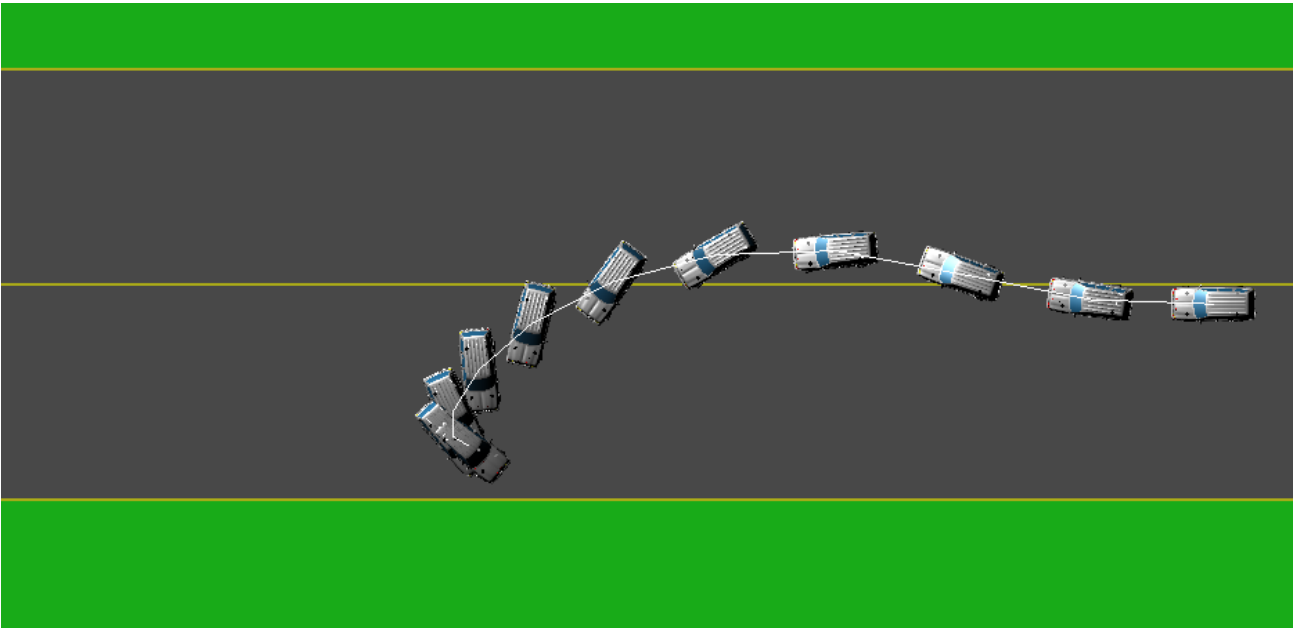

Figure 3 Fishhook maneuver simulation result

**ROLLOVER TREAT WARNING**

Loss of yaw stability is an important active safety consideration in passenger cars. Commercial vehicles have much higher center of gravity location as compared to cars and, thus, tend to roll over during impending loss of yaw stability. Mechanisms to warn against rollover threat and rollover avoidance algorithms are, therefore, crucial for the active safety warning/control systems of commercial vehicles.

Rollover is divided into the two categories of tripped and untripped rollover. Active safety systems are concerned with untripped rollover incidents as they can be avoided by control even though they constitute the smaller percentage of rollover accidents. Tripped rollover cannot be avoided by present active rollover avoidance algorithms and current rollover threat indication/warning systems cannot warn drivers of the possibility of a tripped rollover. Future active safety systems utilizing road sensing through cameras, radar sensors, GPS and map information may detect the possibility of a future tripped rollover by detecting the obstacle that can cause tripping threat and intervene beforehand by modifying the vehicle course accordingly.

In order to study rollover propensity of vehicles and to design rollover avoidance methods in a laboratory setting and also for later road tests, typical untripped rollover creating maneuvers need to be known. The National Highway Traffic and Safety Association NHTSA in the US develops and rates such maneuvers for use in rollover propensity rating of different classes of vehicles. Two typical maneuvers are the J-turn and the fishhook maneuvers. A typical fishhook maneuver simulation is displayed in Figure 3.

The determination of meaningful and accurate rollover threat indicators is an important first step in developing rollover avoidance schemes. Several indicators like the Static Stability Factor (SSF), the Stability Margin (SM), the Tilt Table Ratio (TTR), the Side Pull Ratio (SPR), the Rollover Prevention Metric (RPM) and the Critical Sliding Velocity (CSV) have been around for a long time. Some of these indicators or metrics have also been used by NTHSA for a passive classification of rollover propensity of vehicles.

Most of the currently available rollover warning systems are based on static roll stability measures combined with monitoring of roll angle or lateral acceleration. A warning is issued when the pre-determined threshold is exceeded. The threshold values can be tuned for a particular vehicle to reduce the number of false rollover threat warnings. However, such an approach will be inferior in performance as compared to a dynamic rollover threshold indicator. This is due to the fact that the dynamics of the vehicle is neglected in determining the rollover threat in the passive case.

Rollover prevention of heavy commercial vehicles with liquid load that can slosh is an important active safety application in this area. Jiang (2002) has developed a detailed tractor semi-trailer tanker model with liquid sloshing effect. He has proposed lateral acceleration, trailer sprung mass roll angle and the tire load ratio as three suitable candidates for use as a rollover threat index. He has proposed the use of braking in the event of a rollover threat. This is similar to what some other authors have proposed in the presence of a rollover threat. In other related work, Acarman et al (2002) have proposed a rollover prevention method based on anti-rollover braking and Acarman and Özgüner (2003) have applied frequency shaped sliding mode control to rollover prevention of a tanker truck with liquid load.

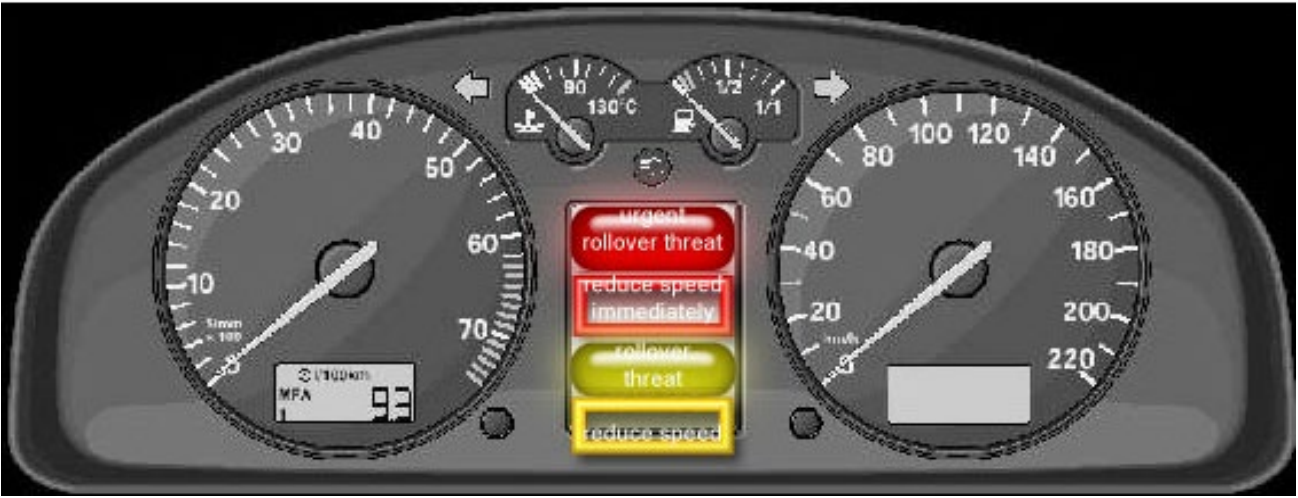

Figure 4 Rollover warning display

There are more recent efforts on the development of dynamic rollover threat indicators that utilize a model of the vehicle for online extrapolation of the vehicle motion into the future until a future rollover event posibility is detected. The Time to Rollover (TTR) metric has been used quite frequently (Chen and Peng 2001, 1999a, 1999b) as a basis for model based rollover threat determination. The critical rollover event is usually defined as tire lift-off or the sprung mass roll angle exceeding a pre-defined value. The TTR value is computed at each time instant by simulating a lower order linear yaw-roll model until the critical rollover event happens at fixed steering wheel position. The closer TTR is to zero, the more critical is the rollover threat. Typical TTR values being greater than 1 sec will allow a good driver enough time to slow down and/or do less aggressive steering. A rollover threat signal will then be sufficient to warn the driver. As commercial vehicles have slower dynamics as compared to SUV's and cars, they also have larger TTR values. Most SUV's have TTR values of 0.5 sec or less (Chen and Peng, 1999a) necessitating the use of a rollover avoidance controller.

Hyun and Langari (2003) have proposed prediction of the LTR (load transfer ratio) of a tractor – semitrailer in real time to indicate rollover threat. They have used a model based method to compute LTR based on available dynamics related measurements. They have proposed estimating variations in some key parameters due to load changes, online, through simplified relations and regression analysis.

The current state-of-the-art in rollover threat indicator calculation is to choose a dynamic model based measure like TTR and to compute it at selected time instants during operation of the vehicle and to display it in a form that is more easily understood by the driver. The display may be in the form of flashing red or yellow indicators (depending on the severity of the situation) and a voice cue that asks the driver to slow down. This is illustrated in Figure 4.

Development of rollover threat warning systems is an active area of research. Dynamic model based methods that take into account the dynamics of the vehicle and mass and mass distribution changes are being developed. The developed algorithms may need a separate ECU or can reside in the ESP ECU if rollover threat indication is bundled with an ESP system. The aim is to reduce the number of false rollover warnings.

**ROLLOVER AVOIDANCE**

A critical TTR value like 1 sec can be used to trigger an active rollover avoidance control system. There are several means of actuation for a rollover avoidance control system. These are: 1) reducing torque and eventually vehicle speed through braking or engine torque reduction, 2) individual wheel braking, 3) active steering, 4) active anti-roll bars, 5) semi-active suspensions and 6) active suspensions. The most direct and easiest way of controlling undesired roll motion is through the use of active suspensions. Since active suspensions have entered the auto industry, the use of active suspension based anti-roll control has become an interesting and promising field. However, reducing vehicle speed through braking and/or engine torque reduction is the simplest and most easily implementable rollover avoidance methodology. Individual wheel braking is also a simple mean of actuation along with active steering in the presence of a steer-by-wire system. However, these two means of actuation also result in yaw motions. There is, thus, the issue of avoiding excessive roll motion while not generating unnecessary yaw motion when these two actuation methods are used. The presence of several actuation methods makes it possible to combine them for better performance.

Rollover avoidance will definitely be a major area of research under preventive and active safety systems. Major improvements, incorporation into ESP controllers and a higher level of availability should be expected in the future.

**LANE KEEPING**

A considerable number of accidents occur due to vehicles that stray from their own lanes. Such undesired lane change may be due to an accidental steering input by the driver or more usually due to a tired or drunk driver. Overworked and therefore sleepy heavy duty vehicle drivers are quite common in some countries. Near ideal highway driving conditions like those on some US highways where the driver drives with set cruise control without having to provide any steering or throttle/brake commands for a long time may also lose concentration while driving even though he is not tired or drunk. A lane departure warning system is quite useful in such situations. The warning system usually works by first sending sound and vision (flashing warning) cues to the driver. The next steps in the case of an unresponding driver are to create a virtual rumble strip by vibrating the steering wheel and to shake the driver seat. In the event that all of these warnings fail, a lane keeping system should take control of the vehicle.

A lane keeping system works by using a camera and a real time image processing algorithm to determine the lane to be followed. This may be complemented by GPS and map data. This information is input as the reference trajectory to a steering controller. Active steering or a steer-by-wire system should be present for implementation. There is a very large literature on intelligent transportation systems (ITS) and also on automatic steering controllers (Schladover, 1978; Schladover et al, 1991; Ackermann and Darenberg, 1990; Unyelioglu et al, 1997; Ackermann et al, 1995; Aksun Güvenç and Güvenç, 2002d). The availability of this wealth of earlier successful results makes it easier to develop reliable automatic lane keeping systems.

**INTELLIGENT TRANSPORTATION SYSTEMS**

The reader should have noted by now that developments in preventive and active safety systems are leading towards the possibility of autonomous and semi-autonomous driving. Autonomous driving or Intelligent Transportation Systems (ITS) have been the subject of tremendous research activity in prominent research centers in the US, Japan and Europe for several decades. The current result is several highway and test track demonstrations of platoons of vehicles or combinations of vehicles on different tracks that drive automatically without driver intervention (Özgüner, 2003). Obviously the roadmap of future developments in the automotive area is taking the route to autonomously driven vehicles. There have also been related major advances in inter-vehicle communications. The trend right now is complementing inter-vehicle communication with information showers from the road to the vehicle (Fujise, 2004). The future possibilities that ITS and information technology offer are very valuable and create a large number of possibilities. However, for widespread acceptance and use of ITS systems, several important issues have to be solved. The most important one is being able to demonstrate ITS technology in actual road conditions and also in cases involving non-ITS vehicles. The ITS control systems have to have a large level of fault tolerance built into them and should be equipped with most of the preventive and active safety systems described here to be ready for unexpected situations. The first commercial use of ITS technology is expected in truck convoys where one monitoring driver will be able to drive a convoy of trucks. Work on off-road use of ITS technology has also started through races like the DARPA Challenge in the US.

**COLLISION AVOIDANCE**

The trend towards autonomous driving necessitates the capability of determining and avoiding obstacles in the desired path of the vehicle. Work on incorporating collision avoidance technology into active vehicle safety control is relatively new (see for ex. Ferrara, 2004) and benefits from the well established theoretical and practical results on obstacle avoidance methods for mobile robots. Road vehicles, however, are much faster than mobile robots that can stop and wait to take the required obstacle avoidance action, if necessary. In both cases moving obstacles are harder to avoid.

Collision avoidance can be broken down into two categories. The first one is determination of a non-collision path between obstacles whose locations have been determined. The second case is the sudden and unexpected detection of an obstacle with imminent collision. The collision avoidance algorithm should urgently perform an evasive maneuver in that case. This is a difficult situation to handle as the evasive maneuver should avoid other possible collisions and loss of lateral or roll stability of the vehicle.

Collision threat warning and indication can be viewed as a prerequisite to designing a collision avoidance system. Collision indicators are available commercially and suffer from false warnings. The use of inter vehicle communication of the brake signal has been proposed for use in a collision indicator based on a detailed study involving road tests (Peng, 2004). A dynamic model based collision indicator metric can be defined and calculated in real time in a manner similar to the time to rollover metric discussed previously. This dynamic TTC (time to collision) metric needs to involve the dynamics of the vehicle that might be collided with along with the original vehicle's dynamics, for accuracy.

Collision indication and collision avoidance technology are very important but still ripe areas of preventive and active safety systems where much more progress needs to be made.

**ADAPTIVE CRUISE CONTROL**

Adaptive cruise control (ACC) is the next stage after cruise control systems. The system uses a radar sensor to detect the vehicle being followed and adapts to reductions in speed of that vehicle. If the vehicle being followed increases its speed above the set limit, it is not followed, resulting in ordinary cruise control at set speed. The ACC system should also differentiate between vehicles on other lanes and not try to follow them. Operation on curved roads should be paid close attention. The controller should not exhibit unpredictable or sudden jumps in desired velocity if the lead vehicle goes out of radar view in a curved road. The simulator in Figure 5 was used in our lab to test ACC algorithms in a two vehicle, two driver scenario (Küçükçalık et al, 2004).

ACC technology has advanced to the level that it is available commercially. Improvements to the control strategies used still need to be carried out. If the pursuit distance drops to about 30 m, one, then, starts talking about a stop-and-go system. This is more difficult as the dynamics of slowing down and speeding become more important. A higher level of demand for ACC and stop and go systems are expected in the future. In cities like İstanbul where stop and go traffic actually means bumper to bumper operation, the development of a true stop and go assistant will be useful. This is a difficult problem as it involves a multiple vehicle scenario, transient engine and vehicle dynamics and collision avoidance simultaneously. Note that this problem becomes relatively easier to implement in hybrid electric vehicles which have the luxury of 1) electric motor shutdown during stoping and 2) smooth and easier to control transient operation during start of following.

**HIL AND SIMULATOR TESTS**

Preventive and active safety applications involve a lot of coding in Electronic Control Units (ECU). The main control algorithm is usually a small part of the code. The larger part of the code is devoted to error checking and what to do in the presence of an error like a sensor malfunction. Preventive and active safety code needs to be checked thoroughly before being approved for vehicle use. While road tests are an important element of this code checking procedure, their use is not the most efficient or exhaustive method of testing. It is much more efficient to connect the related preventive and active safety ECU to a hardware-in-the-loop (HiL) system where the rest of the vehicle dynamics are solved in software in real time in a fast processor in a HiL computer that sends the calculated sensor signals to the ECU. This approach makes very fast and efficient testing available and allows the developer the ability of programming failure scenarios that may not occur or may not be possible in road tests. If the test setup allows real time human driver input, then the simulations are also called human-in the-loop.

The human and hardware in the loop simulation setups used for checking preventive and active safety control system code in the Automotive Controls Research Group at İstanbul Technical University are shown in Figures 5 and 6. The mechanical base of the simulator in Figure 5 is home built. The seat was donated by Ford/Otosan. The setup was built in an undergraduate student design project (see Güvenç and Aksun Güvenç, 2004). A game type steering wheel with force feedback and gas and brake pedals are used for driver input. Two personal computers are used for the main vehicle dynamics simulation. A faster PC with windows is used for the real time animations while a slower PC connected via Ethernet cable and running under xPC® Target runs the vehicle dynamics model computations. The preventive and active safety control algorithms are also simulated in this second PC. It is possible to add two drivers into the loop in this setup by adding a second steering wheel, gas/brake pedal set. The vehicle models are formed in Simulink using the in house Automotive Blockset developed by the Automotive Controls Research Group at İstanbul Technical University.

The mechanical base of the second driving simulator shown in Figure 6 was donated by Meteksan Corp. which builds its driving education trainer using this hardware. The system has real steering wheel and clutch, brake and gas pedals. Rotary potentiometers are used to measure steering wheel and pedal positions. A PC with a dSpace DS1103 card with two processors including a dedicated DSP processor is used for the vehicle model simulation and animation computations. The preventive and active safety code resides in a dSpace ECU. Both the DS1103 and the ECU run in real time. Code is first created in Simulink® and then changed automatically into real time executable code. The ECU and DS1103 card can communicate with each other via the CAN bus or via direct analog I/O connections.

The setup explained above and shown in Figure 6 is being used for testing ESP, rollover threat indicator and rollover avoidance algorithms. It will also be used in a project on safe driving based on vocal and vision based monitoring of the driver. This project has been proposed to the Turkish National Planning Association by a consortium comprising an automotive research entity, three universities, and three major automotive companies in Turkey. One of the main aims of the project is to use driver sound and vision information to determine if the driver is tired or drunk. If it is determined that the driver is not in a condition to drive, the active safety system will enter the picture by either not allowing the vehicle to be driven or in the case of an on road condition by assisting the driving impaired driver in lane keeping and collision avoidance.

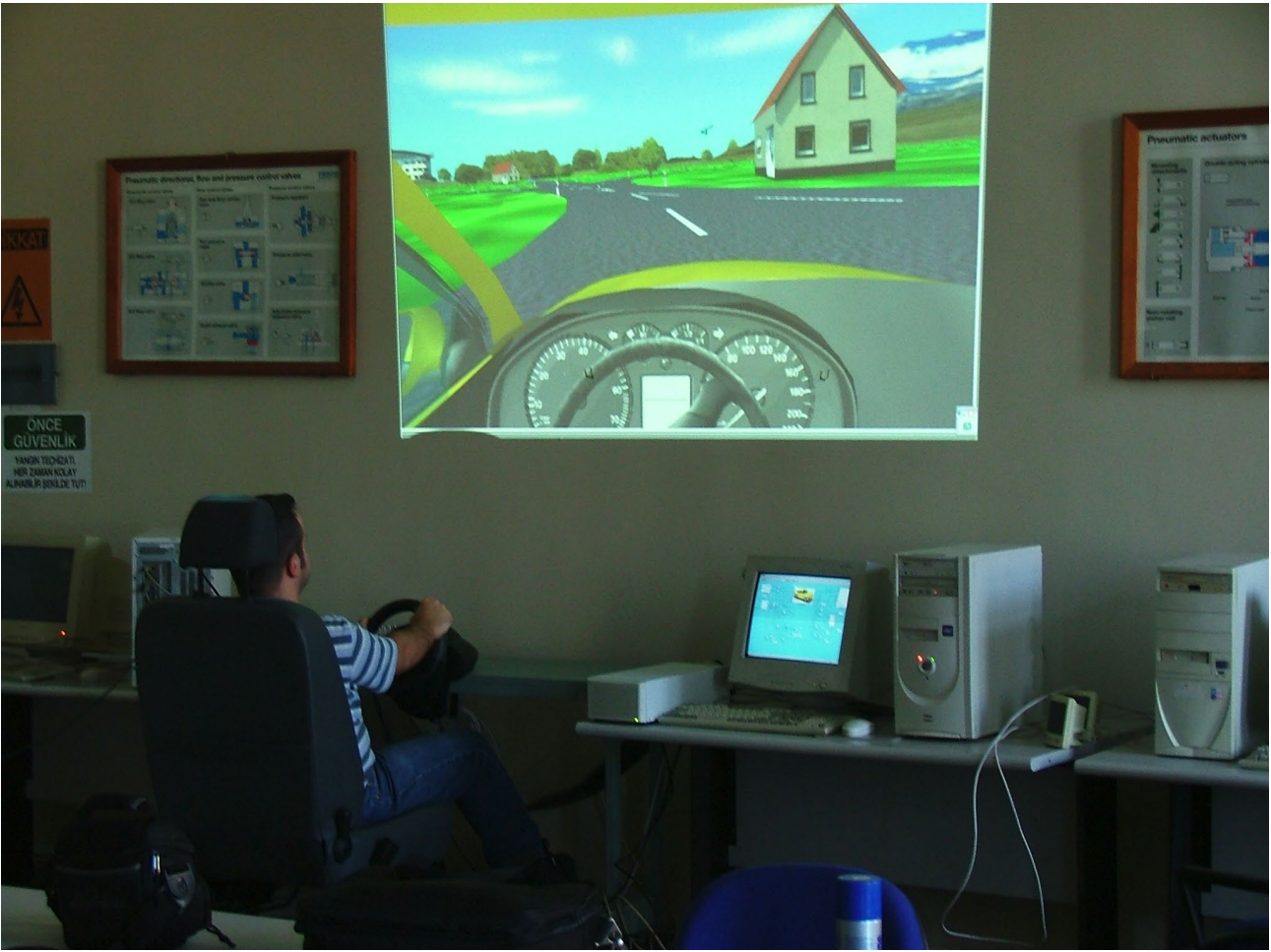

Figure 5 First HiL simulator

Note that human-in-the loop simulation also forms the basis of a simple driving simulator. Subjective and objective driver evaluations of new preventive and active safety systems can be measured in a simulator setting. More advanced simulators usually have 360º of visual cueing capability and a motion platform to impart the vehicle dynamics effects like acceleration and deceleration more accurately to the driver. Advanced simulators can be extremely expensive and are useful for concentrating on certain chosen aspects of the safety system. The ultimate test, of course, is to use a track built for active safety tests and to do extensive road tests.

## THE AUTOMOTIVE CONTROLS RESEARCH GROUP AT İTU

This section first introduces the Automotive Controls Research Group, headed by the author, in the Department of Mechanical Engineering at İstanbul Technical University. The work of this group in preventive and active safety applications is briefly presented in the rest of the section.

The controls related work of the group began in the form of undergraduate student projects and graduate thesis work advised by the author and his colleagues starting in 1997 with the formation of a very small research lab by them in the Department of Mechanical Engineering at İstanbul Technical University. This small research lab grew in the year 1998 with small funds from the university and Tubitak (the equivalent of NSF in Turkey) and a large grant from the National Planning Association of Turkey. This large grant resulted in the formation of our Automation Lab. Due to the presence of large automotive companies in the nearby area which hire a large number of our graduates, part of the focus in the lab turned towards the automotive sector in 1999. By the year 2001, the Automotive Controls Research Group was formed. This group soon attracted its own funding for research, resulting in its own research infrastructure housed in a separate lab. The Automotive Controls Research group now comprises of several faculty members in the Mechanical Engineering Department and has both national and international research collaboration with prominent research labs in automotive control and mechatronics. We have research colleagues in four different universities in İstanbul. One of the aims of our research group is to become a Center of Excellence, coordinating Automotive Controls and Automotive Mechatronics efforts in Turkey. In line with this aim, our research group currently coordinates the Active and Passive Safety Working Group of the Tübitak Automotive Consortium effort for FP6 project participation by Turkey and also the Working Group on Active Safety Systems of the IEEE Technical Committee on Automotive Control.

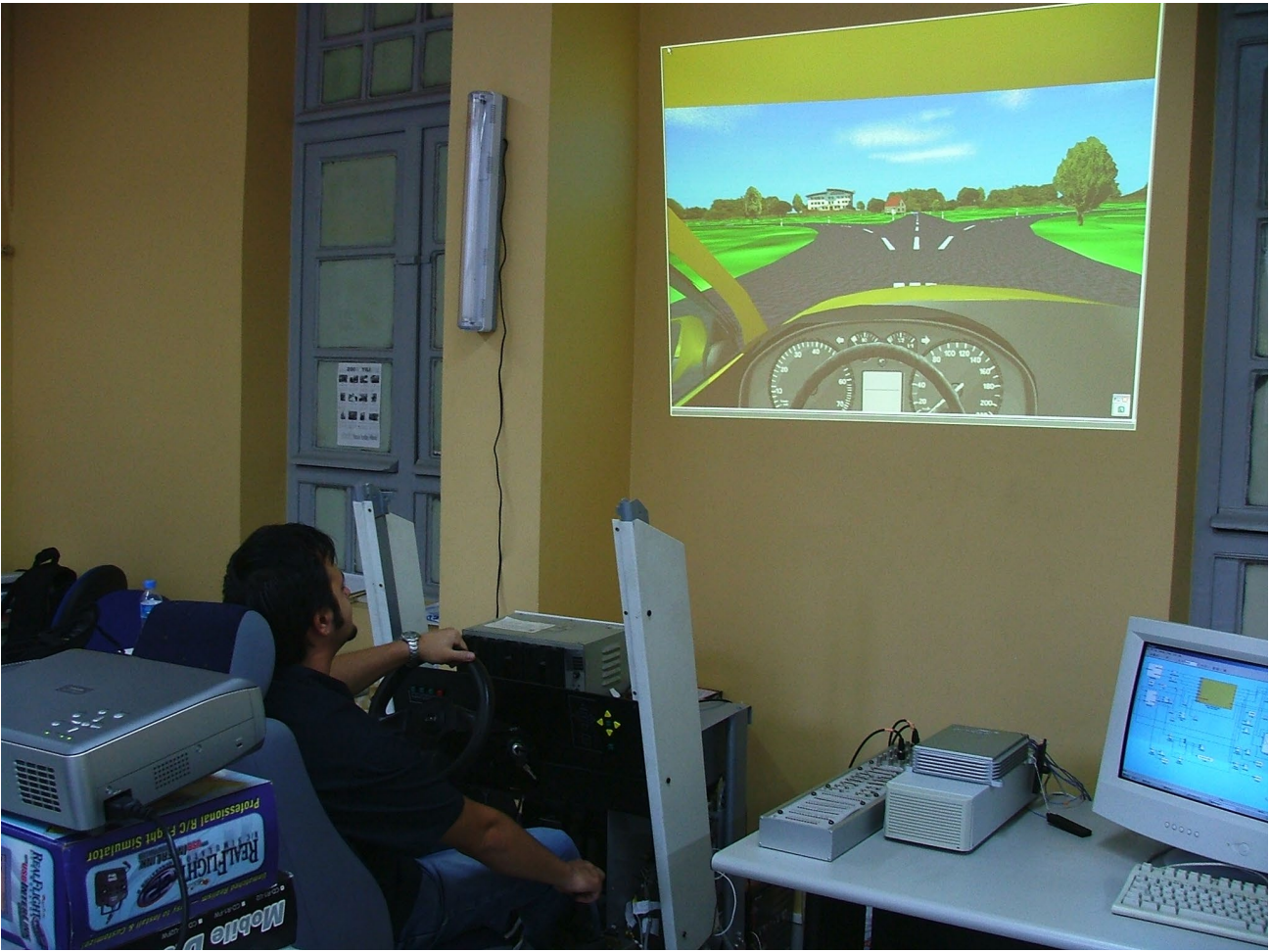

Figure 6 Second HiL Simulator

The activities of the Automotive Controls Research Group related to Preventive and Active Safety include: steering, individual wheel braking and combined actuation ESP control; rollover warning systems; rollover avoidance algorithms; adaptive cruise control; driveline modeling and validation; hardware and human in the loop simulation; real time simulation capable vehicle dynamics model development; suspension parameter optimization; active suspension control and simulator development. The research group is also experienced in engine ECU hardware in the loop test system development in a variety of software and hardware platforms. Some of the future work of the research group will include: hybrid electric vehicle modeling and power management control; lane keeping control systems and collision avoidance.

**ROAD MAP OF EXPECTED FUTURE DEVELOPMENTS AND RECOMMENDATIONS**

A road map of expected future developments in preventive and active safety applications is shown in Figure 7. This road map is an extrapolation based on the current knowledge of the author in this area. Like any extrapolation, it is bound to have errors. Note that such road maps need to be updated regularly in accordance with the actual developments in the field.

The road map shows us that there is a trend towards a fully mechatronic road vehicle with powerful inter-vehicle and road-vehicle communication and autonomous driving capability. The road vehicle of the future will be packed with sensors and electromechanical actuators and will use by-wire technology. The management and control algorithms in these ECUs will be crucial in the future development of road vehicles. Preventive and active safety systems resident in those ECUs will, thus, become very important. The main idea will be for the driver to enter just the destination, and the rest including optimal route planning, will be taken care of by the vehicle. Preventive and active safety systems will be an important background component in making such autonomous driving safe.

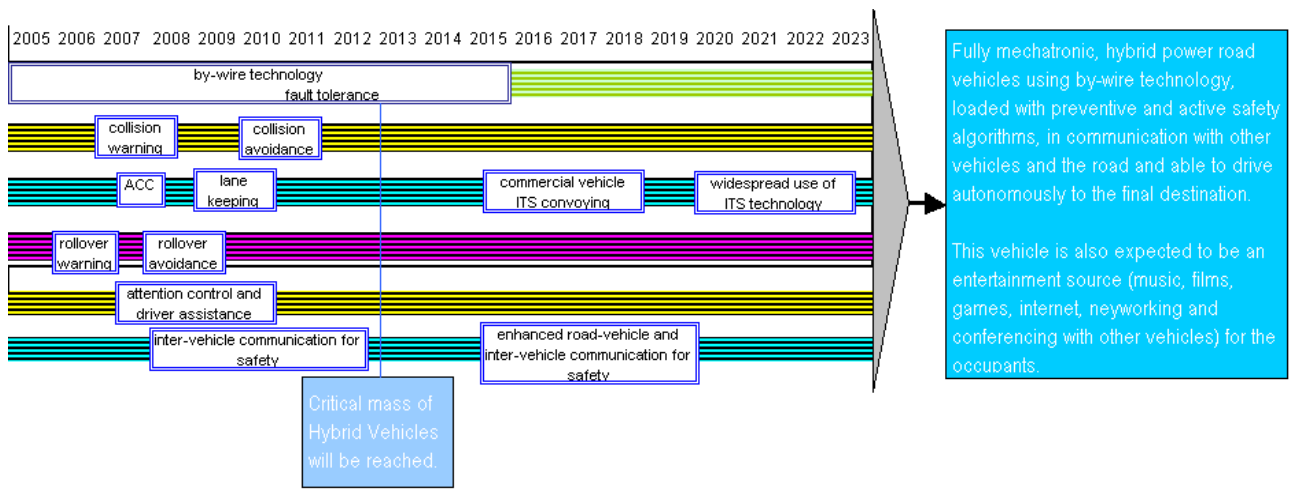

Figure 7 Road map of expected developments in preventive and active safety

Due to the importance of preventive and active safety systems, there are proportionally large research efforts in Europe, the US and Japan. Preventive and active safety research is relatively new in Turkey. This area of research needs to receive more attention and support in Turkey in the future.

Note that only key preventive and active safety systems were concentrated upon in this paper. There are many other useful driver assist technologies like hill hold assistant, parking assistant and enhanced night vision which will also be important components of future vehicles.

**CONCLUSION**

Several key preventive and active safety systems for road vehicles were presented in this paper. Some results from the author's lab were also presented in certain sections. A road map of expected developments in this area was given and several recommendations were made.

It is interesting to note that most of the future predictions made in Figure 7 in 2004 have become a reality in 2020. As predicted in 2004, most of the preventive and active safety methods presented in this paper have evolved into connected and autonomous driving technology in 2020 with most series production vehicles having some from of drive-by-wire capability, many sensors, powerful in-vehicle computing platforms and communication modems. Some research that was reported after 2004 is of interest to the main idea of this paper and is reported next as future work that has taken place. Hybrid electric vehicle have found more widespread use (Boyali and Guvenc, 2010). The drive towards fully electric vehicles has taken place. However, yaw stability control (and traction control) for electric vehicles need a different approach as the electric motor is used for traction instead of the internal combustion engine and as regenerative braking also has to be taken into account (Emirler et al, 2015; Kahraman et al, 2020). Coordinated longitudinal and lateral motion control of electric vehicles has been investigated by Zhou et al (2018). Hardware-in-the-loop simulation methods have been used to accelerate development times and for evaluation purposes (Oncu et al, 2007; Gelbal et al, 2017; Karaman and Guvenc, 2020).

Robust control methods are being used to handle uncertainty (Aksun Guvenc and Guvenc, 2003; Guvenc et al, 2017; Emirler et al, 2014; Emirler et al, 2015; Wang et al, 2018; Wang and Guvenc, 2020). Disturbance observer based methods (Guvenc and Srinivasan; 1994, 1995; Aksun Guvenc and Guvenc, 2001) and repetitive control methods (Orun et al, 2009; Necipoglu et al, 2011a, 2011b; Demirel and Guvenc, 2010; Aksun-Guvenc and Guvenc, 2006) can be taken from other applications and adapted to automotive control in the future.

**ACKNOWLEDGMENTS**

The author would first like to acknowledge the various government and industry sources that have in one way or another supported the research activities of the Automotive Controls Research Group in the Department of Mechanical Engineering at İstanbul Technical University. These sources of support are İstanbul Technical University, Turkish National

Planning Association (DPT), the Scientific and Technical Research Council of Turkey (Tübitak), Ford/Otosan, Meteksan, Tofaş. The author would like to thank the members of the Automotive Control Research Group for their help. The kind invitation of Ford/Otosan to present this material at ICAT 2004 is also acknowledged.